\documentclass[12pt]{article}
\usepackage{amssymb}
\usepackage{graphicx}
\begin{document}
\newcommand{\beq}{\begin{equation}}
\newcommand{\eeq}{\end{equation}}
\newcommand{\beqa}{\begin{eqnarray}}
\newcommand{\eeqa}{\end{eqnarray}}
\newcommand{\beqar}{\begin{eqnarray*}}
\newcommand{\eeqar}{\end{eqnarray*}}
\newcommand{\al}{\alpha}
\newcommand{\be}{\beta}
\newcommand{\del}{\delta}
\newcommand{\D}{\Delta}
\newcommand{\eps}{\epsilon}
\newcommand{\ga}{\gamma}
\newcommand{\Ga}{\Gamma}
\newcommand{\ka}{\kappa}
\newcommand{\nn}{\nonumber}
\newcommand{\inn}{\!\cdot\!}
\newcommand{\h}{\eta}
\newcommand{\ii}{\iota}
\newcommand{\kk}{\varphi}
\newcommand\F{{}_3F_2}
\newcommand{\la}{\lambda}
\newcommand{\La}{\Lambda}
\newcommand{\na}{\prt}
\newcommand{\Om}{\Omega}
\newcommand{\om}{\omega}
\newcommand{\p}{\Phi}
\newcommand{\sig}{\sigma}
\renewcommand{\t}{\theta}
\newcommand{\z}{\zeta}
\newcommand{\ssc}{\scriptscriptstyle}
\newcommand{\eg}{{\it e.g.,}\ }
\newcommand{\ie}{{\it i.e.,}\ }
\newcommand{\labell}[1]{\label{#1}} 
\newcommand{\reef}[1]{(\ref{#1})}
\newcommand\prt{\partial}
\newcommand\veps{\varepsilon}
\newcommand{\pol}{\varepsilon}
\newcommand\vp{\varphi}
\newcommand\ls{\ell_s}
\newcommand\cF{{\cal F}}
\newcommand\cA{{\cal A}}
\newcommand\cS{{\cal S}}
\newcommand\cT{{\cal T}}
\newcommand\cV{{\cal V}}
\newcommand\cL{{\cal L}}
\newcommand\cM{{\cal M}}
\newcommand\cN{{\cal N}}
\newcommand\cG{{\cal G}}
\newcommand\cK{{\cal K}}
\newcommand\cH{{\cal H}}
\newcommand\cI{{\cal I}}
\newcommand\cJ{{\cal J}}
\newcommand\cl{{\iota}}
\newcommand\cP{{\cal P}}
\newcommand\cQ{{\cal Q}}
\newcommand\cg{{\tilde {{\cal G}}}}
\newcommand\cR{{\cal R}}
\newcommand\cB{{\cal B}}
\newcommand\cO{{\cal O}}
\newcommand\tcO{{\tilde {{\cal O}}}}
\newcommand\bz{\bar{z}}
\newcommand\bb{\bar{b}}
\newcommand\ba{\bar{a}}
\newcommand\bg{\bar{g}}
\newcommand\bc{\bar{c}}
\newcommand\bw{\bar{w}}
\newcommand\bX{\bar{X}}
\newcommand\bK{\bar{K}}
\newcommand\bA{\bar{A}}
\newcommand\bZ{\bar{Z}}
\newcommand\bxi{\bar{\xi}}
\newcommand\bphi{\bar{\phi}}
\newcommand\bpsi{\bar{\psi}}
\newcommand\bprt{\bar{\prt}}
\newcommand\bet{\bar{\eta}}
\newcommand\btau{\bar{\tau}}
\newcommand\hF{\hat{F}}
\newcommand\hA{\hat{A}}
\newcommand\hT{\hat{T}}
\newcommand\htau{\hat{\tau}}
\newcommand\hD{\hat{D}}
\newcommand\hf{\hat{f}}
\newcommand\hK{\hat{K}}
\newcommand\hg{\hat{g}}
\newcommand\hp{\hat{\Phi}}
\newcommand\hi{\hat{i}}
\newcommand\ha{\hat{a}}
\newcommand\hb{\hat{b}}
\newcommand\hQ{\hat{Q}}
\newcommand\hP{\hat{\Phi}}
\newcommand\hS{\hat{S}}
\newcommand\hX{\hat{X}}
\newcommand\tL{\tilde{\cal L}}
\newcommand\hL{\hat{\cal L}}
\newcommand\tG{{\tilde G}}
\newcommand\tg{{\tilde g}}
\newcommand\tphi{{\widetilde \Phi}}
\newcommand\tPhi{{\widetilde \Phi}}
\newcommand\te{{\tilde e}}
\newcommand\tk{{\tilde k}}
\newcommand\tf{{\tilde f}}
\newcommand\ta{{\tilde a}}
\newcommand\tb{{\tilde b}}
\newcommand\tc{{\tilde c}}
\newcommand\td{{\tilde d}}
\newcommand\tm{{\tilde m}}
\newcommand\tmu{{\tilde \mu}}
\newcommand\tnu{{\tilde \nu}}
\newcommand\talpha{{\tilde \alpha}}
\newcommand\tbeta{{\tilde \beta}}
\newcommand\trho{{\tilde \rho}}
 \newcommand\tR{{\tilde R}}
\newcommand\teta{{\tilde \eta}}
\newcommand\tF{{\widetilde F}}
\newcommand\tK{{\tilde K}}
\newcommand\tE{{\widetilde E}}
\newcommand\tpsi{{\tilde \psi}}
\newcommand\tX{{\widetilde X}}
\newcommand\tD{{\widetilde D}}
\newcommand\tO{{\widetilde O}}
\newcommand\tS{{\tilde S}}
\newcommand\tB{{\tilde B}}
\newcommand\tA{{\widetilde A}}
\newcommand\tT{{\widetilde T}}
\newcommand\tC{{\widetilde C}}
\newcommand\tV{{\widetilde V}}
\newcommand\thF{{\widetilde {\hat {F}}}}
\newcommand\Tr{{\rm Tr}}
\newcommand\tr{{\rm tr}}
\newcommand\STr{{\rm STr}}
\newcommand\hR{\hat{R}}
\newcommand\M[2]{M^{#1}{}_{#2}}
\newcommand\MZ{\mathbb{Z}}
\newcommand\MR{\mathbb{R}}
\newcommand\bS{\textbf{ S}}
\newcommand\bI{\textbf{ I}}
\newcommand\bJ{\textbf{ J}}

\begin{titlepage}
\begin{center}

\vskip 2 cm
{\LARGE \bf  
Are Genus Corrections in Effective Actions  \\  \vskip 0.25 cm Invariant Under Buscher Rules?
 }\\
\vskip 1.25 cm
 Mohammad R. Garousi\footnote{garousi@um.ac.ir}

\vskip 1 cm
{{\it Department of Physics, Faculty of Science, Ferdowsi University of Mashhad\\}{\it P.O. Box 1436, Mashhad, Iran}\\}
\vskip .1 cm
 \end{center}

\begin{abstract}

It is well-established that the dimensional reduction of the classical effective action of string theory at any order of $\alpha'$ on a circle of arbitrary radius remains invariant under the higher-derivative extension of Buscher transformations. In this study, we extend this symmetry to higher-genus levels. By leveraging the validity of Buscher rules for any genus of the world-sheet, we find that the measure of the effective action remains invariant only when reduced on a self-dual circle. Our findings indicate that the invariance of the Lagrangian density under the higher-derivative and higher-genus extension of the corresponding restricted Buscher rules does not yield the one-loop effective action at order $\alpha'^3$ as derived by the S-matrix method. This result aligns with the general belief that quantum gravity has no global symmetry.

\end{abstract}

\end{titlepage}


The spacetime effective action in string theory features a double expansion in terms of the world-sheet genus $ g$ and the spacetime derivative parameter $ \alpha'$. Various methods are available to determine the $\alpha'$-expansion, including the non-linear sigma model \cite{Tseytlin:1988rr},  T-duality \cite{Garousi:2017fbe}, supersymmetry \cite{Ozkan:2024euj}, and S-matrix method \cite{Gross:1986iv,Gross:1986mw}.
The supersymmetry method, applicable exclusively to superstring theory, leverages spacetime supersymmetry to derive the effective action. In contrast, the  non-linear sigma model, and T-duality methods utilize the conformal symmetry of the world-sheet—a fundamental symmetry present in all string theories, which is considered more fundamental than supersymmetry.

In the non-linear sigma model, the $\alpha'$-expansion of the equations of motion at a given order of $g$ can be determined by ensuring that the 2-dimensional non-linear sigma model remains conformally invariant up to that order. By employing 2-dimensional field theory techniques, one can compute the beta functions of the field theory up to a given order of $ g$. Setting the sum of these beta functions to zero, $\sum_{n=0}^g \beta_n = 0$, allows for the derivation of the equations of motion \cite{Callan:1986bc}, from which the effective action can be obtained.
The beta function at each order of $ g $ has its own $ \alpha' $-expansion, which is related to loop calculations in the 2-dimensional field theory. Specifically, the beta function at order $ \alpha'^m$ corresponds to $(m+1)$-loop calculations. 
 This method has been used at the sphere level ($\beta_0 = 0$) to determine the gravity couplings up to order $\alpha'^3$ \cite{Grisaru:1986kw,Grisaru:1986vi}, and at the torus level ($\beta_0 + \beta_1 = 0$) to find the cosmological constant of the bosonic string theory \cite{Fischler:1986tb}.


The conformal symmetry of the world-sheet theory implies that the non-linear sigma model in two different spacetime backgrounds, each with its own isometries, are related by Buscher transformations \cite{Buscher:1987sk,Rocek:1991ps}, which are independent of the genus of the world-sheet \cite{Hamidi:1986vh}. Consequently, this conformal symmetry suggests that the spacetime effective action of string theory may be symmetric under an extension of the Buscher transformations, which receive $\alpha'$- and $g$-corrections.
This symmetry has been applied at the sphere level to determine NS-NS couplings up to order $\alpha'^3$ \cite{Garousi:2019wgz,Garousi:2019mca,Garousi:2020gio}, which are consistent with sphere-level S-matrix calculations. The corresponding Buscher transformations do receive higher-derivative corrections \cite{Garousi:2019wgz}. It has been proposed in \cite{Liu:2013dna} that certain NS-NS couplings at order $\alpha'^3$ in type II theory, derived from one-loop S-matrix calculations, transform covariantly under Buscher transformations. In this paper, we examine the possibility that the sphere-level symmetry can be extended to higher genus as well, by including genus and higher-derivative corrections to the Buscher rules.

Consider the simple case where the background fields in the sigma model are independent of the Killing coordinate $y$. In this scenario, the two background NS-NS fields $\Phi, B_{\alpha\beta}, G_{\alpha\beta}$ and $\Phi', B'_{\alpha\beta}, G'_{\alpha\beta}$ are related by the following Buscher rules \cite{Buscher:1987sk}:
\beqa
e^{2\Phi'}=\frac{e^{2\Phi}}{G_{yy}}\,\,\,;\,\,\,G'_{\mu y}=\frac{B_{\mu y}}{G_{yy}}&;&
G'_{\mu\nu}=G_{\mu\nu}-\frac{G_{\mu y}G_{\nu y}-B_{\mu y}B_{\nu y}}{G_{yy}}\nonumber\\
G'_{yy}=\frac{1}{G_{yy}}\,\,\,;\,\,\,B'_{\mu y}=\frac{G_{\mu y}}{G_{yy}}&;&
B'_{\mu\nu}=B_{\mu\nu}-\frac{B_{\mu y}G_{\nu y}-G_{\mu y}B_{\nu y}}{G_{yy}}\labell{tree}
\eeqa
where $\mu,\nu$ denote any   direction other than $y$. In above transformation the metric is  in the string frame. These global transformations form a $\MZ_2$ group. The aforementioned transformation is valid for any world-sheet genus ( see \eg \cite{Alvarez:1994dn,Giveon:1994fu}). The measure of the tree-level effective action, given by $e^{-2\Phi}\sqrt{-\det G_{\alpha\beta}}$, is invariant under this transformation. The tree-level Lagrangian density, at any order of $\alpha'$, is also  invariant under the higher-derivative extension of the aforementioned transformations. The higher-derivative extension of the Buscher rules  also form a $\mathbb{Z}_2$ group \cite{Garousi:2019wgz,Garousi:2019mca,Garousi:2020gio}. 

At higher genus, the leading $\alpha'$ order term of the effective action is the cosmological constant term. The cosmological constant of the bosonic string theory at $g = 1$, which is non-zero (see \eg \cite{Polchinski:1985zf,Peskin:1987rz}), is $\sqrt{-\det G_{\alpha\beta}}\,C$ and may be invariant under the Buscher transformation. More generally, at the $g$-loop level, the measure of the effective action of oriented closed string is $e^{2(g-1)\Phi}\sqrt{-\det G_{\alpha\beta}}$, which may also be invariant under the  Buscher transformation. However, this measure is not invariant under the classical Buscher rules \reef{tree}. Given that the leading-order T-duality transformations hold true at each genus order, it can be observed that the measure remains invariant only when the Killing circle is self-dual, i.e., $G_{yy} = 1$ for $g > 0$. Therefore, if one considers the following genus expansion for the effective action of oriented closed string theory:
\beqa
\bS&\sim& \sum_{g=0}^{\infty}\int d^Dx \,e^{2(g-1)\Phi}\sqrt{-\det G_{\alpha\beta}}\,\cL_g\labell{Sg}
\eeqa
one may expect the Lagrangian density $\cL_g$  to be invariant under the following restricted Buscher transformation for $g>0$:
\beqa
\Phi'=\Phi\,\,\,;\,\,\,
G'_{\mu y}=B_{\mu y}&;&
G'_{\mu\nu}=G_{\mu\nu}-(G_{\mu y}G_{\nu y}-B_{\mu y}B_{\nu y})\nonumber\\
B'_{\mu y}=G_{\mu y}&;&
B'_{\mu\nu}=B_{\mu\nu}-(B_{\mu y}G_{\nu y}-G_{\mu y}B_{\nu y})\labell{oneloop}
\eeqa
The Lagrangian density $\mathcal{L}_g$ at genus $g$ possesses its own $\alpha'$-expansion. At the leading order of $\alpha'$, it may be invariant under the aforementioned restricted Buscher transformations. Furthermore, all higher-order terms in $\alpha'$ may be invariant under the higher-derivative and higher-genus extensions of these transformations.

Considering that the overall dilaton factor in the effective action depends on the genus of the world-sheet, it is observed that the higher-derivative extension of the restricted Buscher transformations \reef{oneloop} should undergo a specific genus expansion that respects the Euler characteristic of the oriented world-sheet. Specifically, when the higher-genus expansion is incorporated into the effective action, it should yield the overall dilaton factor $e^{2(n-1)\Phi}$ for $n > 0$. This principle should similarly apply to higher-genus field redefinitions. We investigate whether the effective action of string theory at one-loop order remains invariant under the higher-derivative extension of the aforementioned restricted Buscher transformations. Additionally, we consider the scenario where these transformations receive a higher-genus expansion.

The extension of the classical Buscher transformation \reef{tree} to the torus $T^{(n)}$ involves replacing $G_{yy}$ with $Q_{ij}$, where $i$ and $j$ are the indices along the torus, and $Q_{ij} = G_{ij} + B_{ij}$. The self-dual restriction $G_{yy} = 1$ for the circular reduction imposes the condition $Q_{ij} = \delta_{ij}$ on the torus reduction. The transformations of the NS-NS fields are then given by:
 \beqa
\Phi'=\Phi\,\,\,;\,\,\,
G'_{\mu i}=B_{\mu i}&;&
G'_{\mu\nu}=G_{\mu\nu}-(G_{\mu i}G_{\nu i}-B_{\mu i}B_{\nu i})\nonumber\\
B'_{\mu i}=G_{\mu i}&;&
B'_{\mu\nu}=B_{\mu\nu}-(B_{\mu i}G_{\nu i}-G_{\mu i}B_{\nu i})\labell{oneloopi}
\eeqa
They indicate that the cosmological reduction of the loop effective action lacks symmetry, in contrast to the classical theory, which possesses  the global $O(d,d)$-symmetry \cite{Sen:1991zi,Hohm:2014sxa}. However, there might be the global $\mathbb{Z}_2$-symmetry for the torus reduction of the effective action  for $n < d$. We examine this symmetry specifically for circular reduction.

A similar question arises regarding the $\mathbb{Z}_2$-symmetry of the effective action of unoriented  type I string theory: Are the genus corrections invariant under the restricted T-duality transformations \reef{oneloop}? The first genus corrections in this case correspond to the disk and projective-plane world-sheets. Therefore, the open-closed string couplings  are intrinsically quantum effects and may be invariant under the restricted T-duality transformations \reef{oneloop}. In this context, the measure, given by $e^{-\Phi}\sqrt{-\det G_{\alpha\beta}}$, is invariant under the Buscher transformations \reef{tree}. However, the Lagrangian density may be invariant solely under the restricted Buscher transformation \reef{oneloop}.
This implies that the open-closed string couplings on the D$p$-brane world-volume which are quantum effect may  be invariant under the following  restricted  Buscher transformations:
\beqa
A_\ta'=A_\ta\,\,\,;\,\,\,\Phi'=\Phi\,\,\,;\,\,\,
G'_{\mu y}=B_{\mu y}&;&
G'_{\mu\nu}=G_{\mu\nu}-(G_{\mu y}G_{\nu y}-B_{\mu y}B_{\nu y})\nonumber\\
X'^\mu=X^\mu\,\,\,;\,\,\,A_y' = X^y\,\,\,;\,\,\,B'_{\mu y}=G_{\mu y}&;&
B'_{\mu\nu}=B_{\mu\nu}-(B_{\mu y}G_{\nu y}-G_{\mu y}B_{\nu y})\labell{oneloopA}
\eeqa
where $A_a$ is the world-volume gauge field, $X^\alpha$  are the world-volume fields that embed the D$_p$-brane into spacetime, and the index $\tilde{a}$ denotes the world-volume directions other than the $y$-direction. In this case, the measure  $e^{-\Phi}\sqrt{-\det P[G_{ab} + B_{ab}] + F_{ab}}$, where the pull-back is $ P[G_{ab}] = \partial_a X^\alpha \partial_b X^\beta G_{\alpha\beta}$, is also invariant under  classical   transformations. However, the Lagrangian density may be invariant solely under the higher-derivative extension of the above quantum  transformation.
In fact, we have performed explicit calculations  at order $\alpha'$ for couplings involving the Riemann curvature, dilaton, the second fundamental form $ \Omega_{ab}{}^\alpha$, and $ F_{ab}$, considering the simple case where $ G_{\mu y} = 0 = B_{\mu y} $. We have attempted to fix the coupling constants of these world-volume couplings by imposing invariance under the higher-derivative extension of the classical Buscher transformation. However, we found that the T-duality constraint yields zero results unless the couplings are required to be invariant under the higher-derivative  extension of the restricted transformation \reef{oneloopA}. Similar calculations for classical couplings involving only the open string case have been performed in \cite{Baradaran-Hosseini:2024xap}. This calculation indicates that the open-closed couplings  for non-zero $G_{\mu y}$ and $B_{\mu y}$ cannot be invariant under the classical Buscher rules. They may, at most, be invariant under  the restricted Buscher transformations \reef{oneloopA}.

Let us examine the proposal that the one-loop effective action of NS-NS fields in the heterotic theory is invariant under the restricted Buscher transformation \reef{oneloop} for non-zero $G_{\mu y}$ and $B_{\mu y}$. Supersymmetry implies that the cosmological constant is zero\footnote{It can be observed that a particular field redefinition can absorb the tree-level effective Lagrangian $e^{-2\Phi}(R + 4(\nabla\Phi)^2 - \frac{1}{12}H^2)$ into the one-loop cosmological constant of the bosonic theory. Under such field redefinitions, there would be no coupling in the tree-level effective action. A similar observation has been made in \cite{Codina:2023fhy} for higher-derivative field redefinitions in the tree-level effective action of non-critical bosonic string theory.
This apparent puzzle, where the field redefinition absorbs the tree-level effective action into the one-loop cosmological constant, is resolved by noting that this field redefinition would also produce other higher-order couplings with overall dilaton factors $e^{-4\Phi}, e^{-6\Phi}, \cdots$, which are inconsistent with the Euler characteristic of the world-sheet. Therefore, such field redefinitions are not permissible.}.
 Consequently, the leading $\alpha'$-order action consists of three independent terms: $R$, $H^2$, and $(\nabla\Phi)^2$. Since the dilaton remains invariant under the T-duality transformation \reef{oneloop}, the T-duality constraint can be imposed in the case of a constant dilaton. Therefore, we focus on the independent couplings for the metric and the $ B$-field only. The invariance of the two couplings, $R$ and $H^2$, under the transformation \reef{oneloop} determines the two coupling constants in terms of a single factor as follows:
\beqa
\bS_1^0&\sim&a\int d^{10}x \sqrt{-G}(R-\frac{1}{12}H^2)\labell{S10}
\eeqa
where $a$ is the overall factor that cannot be fixed by T-duality. The S-matrix method indicates that there are no one-loop corrections to the leading-order classical effective action \cite{Schwarz:1982jn}, hence, $a = 0$.


At order $\alpha'$, there is the freedom to use higher-derivative field redefinitions at each genus. However, since the two-derivative couplings at the one-loop level are zero, there is no such freedom here. Using only the Bianchi identities and integration by parts, we find that the  basis consists of the following 10 couplings:
\beqa
 \bS_1^1&\sim&\alpha'\int d^{10}x \sqrt{-G}\Big[a_{1}   H_{\alpha  }{}^{\delta  \epsilon  } H^{\alpha  
\beta  \gamma  } H_{\beta  \delta  }{}^{\varepsilon  } 
H_{\gamma  \epsilon  \varepsilon  } +   
 a_{2}   H_{\alpha  \beta  }{}^{\delta  } H^{\alpha  
\beta  \gamma  } H_{\gamma  }{}^{\epsilon  \varepsilon  } 
H_{\delta  \epsilon  \varepsilon  } +   
 a_{3}   H_{\alpha  \beta  \gamma  } H^{\alpha  \beta  
\gamma  } H_{\delta  \epsilon  \varepsilon  } H^{\delta  
\epsilon  \varepsilon  }\nn\\&& +   
 a_{7}   H_{\alpha  }{}^{\gamma  \delta  } H_{\beta  
\gamma  \delta  } R^{\alpha  \beta  } +   
 a_{4}   R_{\alpha  \beta  } R^{\alpha  
\beta  } +   
 a_{9}   H_{\alpha  \beta  \gamma  } H^{\alpha  \beta  
\gamma  } R +   a_{5}   R^2 +   
 a_{6}   R_{\alpha  \beta  \gamma  \delta  } 
R^{\alpha  \beta  \gamma  \delta  }\nn\\&& +   
 a_{10}   H_{\alpha  }{}^{\delta  \epsilon  } H^{\alpha  
\beta  \gamma  } R_{\beta  \gamma  \delta  \epsilon  } 
+  a_{8}   \nabla_{\gamma  }H_{\alpha  \beta  \delta  } 
\nabla^{\delta  }H^{\alpha  \beta  \gamma  } \Big]\labell{S1b}
\eeqa
Here, $a_1, \cdots, a_{10}$ represent the 10 coupling constants that may be determined by T-duality.  By imposing the constraint that the circular reduction of these couplings be invariant under the restricted Buscher transformation \reef{oneloop}, we identify the following two T-dual multiplets:
\beqa
 \bS_1^1&\sim&\alpha'\int d^{10}x \sqrt{-G}\Big[    a_{3} ( 12 R-H^2)^2  +   
 a_{10}  \Big( H_{\alpha  }{}^{\gamma  \delta  } H_{\beta  
\gamma  \delta  } R^{\alpha  \beta  }- \frac{1}{4}   
    H_{\alpha  \beta  }{}^{\delta  } H^{\alpha  
\beta  \gamma  } H_{\gamma  }{}^{\epsilon  \varepsilon  } 
H_{\delta  \epsilon  \varepsilon  }\nn\\&& -4   
   R_{\alpha  \beta  } R^{\alpha  
\beta  }  +  
   H_{\alpha  }{}^{\delta  \epsilon  } H^{\alpha  
\beta  \gamma  } R_{\beta  \gamma  \delta  \epsilon  } 
 -     \nabla_{\gamma  }H_{\alpha  \beta  \delta  } 
\nabla^{\delta  }H^{\alpha  \beta  \gamma  }\Big)\Big]\labell{S1b1}
\eeqa
This calculation is similar to the tree-level calculations \cite{Garousi:2019wgz}, where there are no two-derivative couplings at one-loop, i.e., $a = 0$ in \reef{S10}, and hence no two-derivative corrections to the transformation \reef{oneloop}. Since there is no Riemann squared term in the above T-dual multiplets, they can be removed by using the tree-level equations of motion. This result is consistent with the fact that there are no one-loop corrections to the heterotic string theory for NS-NS couplings at order $\alpha'$ \cite{Sakai:1986bi, Ellis:1987dc, Abe:1988cq, Abe:1987ud}. Since it is consistent with the S-matrix, we do not consider the possibility that the transformations \reef{oneloop} receive higher-genus corrections at order $\alpha'$.

At order $\alpha'^2$, by utilizing the Bianchi identities and performing integration by parts, we determine that the basis comprises 70 parity-even and 13 parity-odd couplings. Since there are no 2-derivative and 4-derivative couplings at one-loop order, the T-duality transformation \reef{oneloop} does not receive 2- and 4-derivative corrections. The constraint that the 83 couplings be invariant under the transformation \reef{oneloop} produces four T-dual multiplets. None of them include a Riemann cubed term, and all of them are removable by using the tree-level equations of motion. This result is again consistent with the fact that there are no one-loop corrections for NS-NS couplings at order $\alpha'^2$ in the heterotic theory \cite{Sakai:1986bi, Ellis:1987dc, Abe:1988cq, Abe:1987ud}. 
Since it is consistent with the S-matrix, we do not consider the possibility that the transformations \reef{oneloop} receive higher-genus corrections at order $\alpha'^2$. Producing a zero result for the couplings, even though consistent with the S-matrix calculation, is not enough to verify the proposal that the one-loop effective action is invariant under T-duality. One should produce some non-zero couplings that are derived using the S-matrix method.

 The first nontrivial T-dual multiplet in the heterotic theory, which is not removable by tree-level equations of motion, should appear at order $\alpha'^3$ because the one-loop gravity couplings at this order are non-zero \cite{Sakai:1986bi,Ellis:1987dc,Abe:1988cq,Abe:1987ud}. Moreover, the one-loop corrections to type II theories also begin at this order \cite{Sakai:1986bi}. Since type IIA string theory transforms under T-duality to type IIB string theory, one may expect $\bS_A \rightarrow \bS_B$. Given that the T-duality transformation satisfies the $\mathbb{Z}_2$ group, it is generally expected that the NS-NS part of the effective action in these two theories will have the following structures:

\beqa
\bS_{A}=S_1-S_2 &;& \bS_B=S_1+S_2
\eeqa
where $S_1$ is symmetric under the T-duality transformation, whereas $S_2$ is antisymmetric. At tree-level and at order $\alpha'^3$, $S_2 = 0$, because $S_1$ which is fixed by T-duality produces all 8-derivative NS-NS couplings in type II theories \cite{Garousi:2022ghs}. However, the gravity couplings at the one-loop level indicate that both $S_1$ and $S_2$ are non-zero, \ie  $S_1=t_8 t_8 R^4+\cdots$ and $S_2=\frac{1}{8}\epsilon_{10}\epsilon_{10}R^4+\cdots$. Some of the five-field couplings in $S_1$ and $S_2$ have been found in \cite{Liu:2013dna} through one-loop S-matrix calculations. 

We have performed the calculation at $\alpha'^3$ order for parity-even couplings in the heterotic theory. The basis, which includes all independent terms with more than two fields, contains 937 couplings. Since there are no 2-, 4-, and 6-derivative couplings at one-loop order, we must examine the invariance of these 937 couplings under the T-duality transformation \reef{oneloop}, up to Bianchi identities and total derivative terms in the base space. We have identified 9 T-dual multiplets. However, all of them are removable using the tree-level equations of motion. The calculation is similar to the one considered in \cite{Garousi:2020gio}.

We have also considered the 872 independent couplings found in \cite{Garousi:2020mqn}, where the equations of motion are used, and assumed a constant dilaton. We examine different coupling constants for these couplings in type IIA and type IIB. We impose the condition that the transformation of the circular reduction of type IIA couplings under the T-duality transformation \reef{oneloop} matches the circular reduction of the couplings in type IIB. We found that this constrains all coupling constants in types IIA and IIB to be zero. This is in contrast to the proposal made in \cite{Liu:2013dna} that the one-loop couplings in type IIA transform to type IIB under T-duality.

We have also performed the above calculation in the heterotic theory, assuming that the one-loop T-duality transformation \reef{oneloop} receives genus corrections at order $\alpha'^3$, resulting in some non-zero contributions from the tree-level effective action. In this case, the non-zero couplings that T-duality produces for gravity are exactly $ t_8 t_8 R^4 + \frac{1}{8} \epsilon_{10} \epsilon_{10} R^4 $, which is not the one-loop effective action of heterotic theory at this order \cite{Ellis:1987dc}. Instead, it is the classical effective action of type II theory that has already been found \cite{Garousi:2020gio} by imposing the classical Buscher rules \reef{tree}.

Therefore, 
our calculations indicate that the invariance of the spacetime effective actions or the covariance of the D-brane effective actions in string theory under the derivative extension of the global Buscher transformations is valid only for the classical theory. Given that string theory is a candidate for quantum gravity, this result is consistent with the general belief that quantum gravity has no global symmetry ( see \eg \cite{Harlow:2018jwu}). 



\end{document}